\documentclass[12pt]{iopart}

\usepackage{graphicx}% Include figure files
\usepackage{dcolumn}% Align table columns on decimal point
\usepackage{bm}% bold math
\usepackage{subfigure}
\usepackage{array}
\usepackage{hyperref}
\usepackage{epsfig}
\usepackage{hyperref}
\begin{document}

\title{Magnetic interaction of Co ions near the \{10$\bar{1}$0\} ZnO surface} 

\author{Thomas Archer, Chaitanya Das Pemmaraju and Stefano Sanvito}
\address{School of Physics and CRANN, Trinity College, Dublin 2, Ireland}

\date{\today}
\begin{abstract}

Co-doped ZnO is the prototypical dilute magnetic oxide showing many of the characteristics of 
ferromagnetism. The microscopic origin of the long range order however remains elusive, 
since the conventional mechanisms for the magnetic interaction, such as super-exchange and double 
exchange, fail either at the fundamental or at a quantitative level. Intriguingly, there is a growing evidence 
that defects both in point-like or extended form play a fundamental role in driving the magnetic order. Here 
we explore one of such possibilities by performing {\it ab initio} density functional theory calculations for 
the magnetic interaction of Co ions at or near a ZnO \{10$\bar{1}$0\} surface. We find that extended surface
states can hybridize with the $e$-levels of Co and efficiently mediate the magnetic order, although 
such a mechanism is effective only for ions placed in the first few atomic planes near the surface. 
We also find that the magnetic anisotropy changes at the surface from an hard-axis easy-plane to an
easy axis, with an associated increase of its magnitude. We then conclude that clusters with high densities
of surfacial Co ions may display blocking temperatures much higher than in the bulk.

\end{abstract}

\vspace{2pc}
\submitto{\NJP}

\pagebreak

\section{Introduction}
\label{}

% DMS status
Dilute magnetic semiconductors (DMSs)~\cite{DMS} are a new class of materials in which transition metal (TM) dopants 
(usually Co, Fe or Mn) replace native cations in ordinary semiconductors. The TMs provide localized spins and 
in some cases free carriers and the material often displays evidence for ferromagnetism at remarkably low 
TM concentrations and relative high temperatures. The expectation around DMSs is that they might offer the 
advantages of semiconductors (i.e. an easy to manipulate electronic structure) together with the non-volatility 
of ferromagnets. In short they may be an ideal materials platform for future ``spintronics'' technologies \cite{spin}.

(Ga,Mn)As is by far the most studied among the DMSs, mostly because of its compatibility with present the
GaAs/AlAs technology. However, despite more than a decade of refinements in the synthetic method and
of improvements of the annealing treatments to control the various defects concentrations, the (Ga,Mn)As 
Curie temperature still remains far from exceeding room temperature \cite{Nottingham}, hindering the 
prospect for mainstream applications. These difficulties have stimulated the experimental activity towards 
other host materials and in particular towards oxides based DMSs \cite{jmdcoey} such as ZnO.

The most commonly used dopant in this case is Co and at present the claims for room temperature 
ferromagnetism in ZnO:Co, following the first exciting report from Ueda et al.~\cite{ueda}, are numerous
(for a review see \cite{Das_ZnO} and references therein). The experimental literature however is populated 
with controversial results and almost all possible magnetic phases have been found in samples with the same 
nominal Co concentration and grown under apparently similar conditions. Thus together with ferromagnetism,
also paramagnetism \cite{anisotropy} and a spin-glass behavior  \cite{peng} have been reported. 

Already early experiments have pointed to the extreme sensitivity of the magnetic properties over the sample 
preparation, the post-growth treatment and the various growth conditions. Deposition protocols at controlled 
thermodynamical equilibrium, such as chemical synthesis~\cite{wi,risbud} and molecular beam epitaxy~\cite{chambers} 
generally lead to paramagnetic films, while out of equilibrium schemes such as pulsed laser deposition by large 
produce room temperature magnetism~\cite{venk}. One interesting common feature is that in the cases where 
ferromagnetism is observed the saturation magnetization and the remanence are typically small~\cite{ups,taun,lee}, 
while proving a direct correlation between magnetism and carriers gave conflicting results \cite{behan,Khare}.
More recently, the growth of ultra-high quality ZnO:Co films lacking of magnetism even for large $n$-doping
conditions \cite{Scott1,Scott2}, indicated clearly that the magnetism must be related to the presence of 
crystallographic defects, either in their point-like or macroscopic form. This line of thoughts was used
to critically re-analyse a multitude of published data for ZnO:Mn and to correlate the appearance of magnetism 
to a large surface to volume ratio of the sample grains \cite{straumal}.  

With the conventional mechanisms for the magnetic interaction, such as super- and double-exchange, failing either at the 
qualitative or quantitative level \cite{Ruairi}, most of the theoretical activity in the field has concentrated on first principles
methods, in particular on density functional theory (DFT). Early calculations for ZnO doped with extremely large 
concentrations of Co (25\%) predicted a room temperature ferromagnetic ground state in the absence of additional 
dopants \cite{olddft, olddft2}. These concentrations however are above the percolation threshold for nearest neighbors 
magnetic interaction, so that even a short range mechanism as super-exchange can, in principle, yield to ferromagnetism. 
Below such a threshold, where essentially all the experiments are conducted, one should expect at best only
super-paramagnetic clusters \cite{Das_ZnO}.

Furthermore, it is important to point out that for this rather complicated problem, one should take a particular care 
when choosing the exchange and correlation functional to use in DFT. The standard local density and generalized 
gradient approximations (respectively LDA and GGA) severely under-estimate the native ZnO gap and 
misplace in energy the Co $d$ levels. Both effects contribute to predicting spurious long-range ferromagnetism. Rigorous 
corrections based on energy considerations \cite{Das_Comment,Lany,Priya} improve the description and return a picture 
of antiferromagnetic interaction among nearest neighbors Co ions and little interaction at any other distance \cite{Das_ZnO}. 
Thus also theory points to defects as source of magnetism or at least as a tool to boost either the strength or the range of the 
magnetic interaction. 

For instance uncompensated spins on the surface of secondary CoO phases have been proposed as the origin for the experimentally 
observed  magnetic hysteresis \cite{ditel}. This essentially means to attribute the observed ferromagnetism to super-paramagnetically 
blocked clusters. However, this proposal was also rejected at the quantitative level by DFT calculations. In fact the calculated order 
temperatures of cubic, wurtzite and zinc-blende CoO polymorphs are all well below room temperature~\cite{Tom_CoO}, with finite 
size effects not playing any particular role \cite{Ruairi2}.

Another possibility is represented by complexes involving Co ions and native ZnO intrinsic defects. This was examined in
great detailed by Pemmaraju~et al~\cite{Das_ZnO}, who concluded that only complexes of O vacancies (V$_\mathrm{O}$) and 
substitutional Co ions can couple magnetically to a medium range if additional $n$-type doping is present. Other defect 
combinations can provide remarkably large local magnetic interaction but provide no long range interaction. A phase diagram was then proposed 
where the various magnetic states could be mapped out onto the relative abundance of Co ions and Co-V$_\mathrm{O}$ 
complexes. Although suggestive the presence of V$_\mathrm{O}$ in magnetic samples is often claimed,
it is not clear whether the concentration of Co-V$_\mathrm{O}$ complexes needed for room temperature ferromagnetism
are achievable in reality \cite{VO}. 

In this work we look at yet another possibility, i.e. we investigate whether the magnetism can arise at a ZnO open surface,
namely the non-polar  \{10$\bar{1}$0\} surface. If this is proved true then theory will also converge in establishing a link 
between the macroscopic morphology of samples and their magnetism.

\section{Calculation Details}

Our calculations are all performed by using DFT. The standard local and semi-local exchange and correlation functionals (LDA and GGA) 
are not appropriate for the description of the electronic structure of ZnO and CoO~\cite{Das_ZnO}. We have therefore opted for an 
approximated form of the self-interaction correction scheme (ASIC)~\cite{asic}, which has been shown to correctly reproduce the electronic 
structure of ZnO:Co~\cite{Das_ZnO}, in good agreement with spectroscopical data. The ASIC method is included in a development version of
the local atomic orbital basis set DFT code \textit{Siesta} \cite{SIESTA}. For all our calculations we use the value of $\alpha=\frac{1}{2}$ for 
the ASIC potential scaling parameter. This provides a good description of the electronic structure of strongly correlated insulators and 
semiconductors, away from the purely ionic limit \cite{asic}. 

The electron density and the Hamiltonian for the valence electrons are described by a double-$\zeta$ polarized (DZP) basis set. Core 
electrons are replaced by standard norm-conserving Troullier-Martins' pseudopotentials. The real space integration grid has a 
spacing equivalent to a 800~Ry plane-wave cut-off, while the reciprocal space sampling is performed over a grid with an equivalent real 
space distance of 15~\AA. The forces are calculated from the LDA Ceperley-Alder~\cite{Ceperley80} functional using the ASIC density
\cite{Andrea}, and all structures are relaxed to less than 0.01~eV/\AA.

\section{Un-doped ZnO surface}

The wurtzite structure of bulk ZnO is well described by the ASIC method: both the in-plane, \textbf{a}, and the out of plane, \textbf{c}, lattice 
parameters are slightly contracted with respect to experiments. In particular they are calculated to be respectively 3.209~\AA\ and 
5.180~\AA\, against the reported values of 3.252~\AA\ and 5.313~\AA. The distortion parameter $u$ after relaxation remains essentially 
identical to its experimental value. 
\begin{figure}[ht]
\centering{
\includegraphics[width=0.99\linewidth]{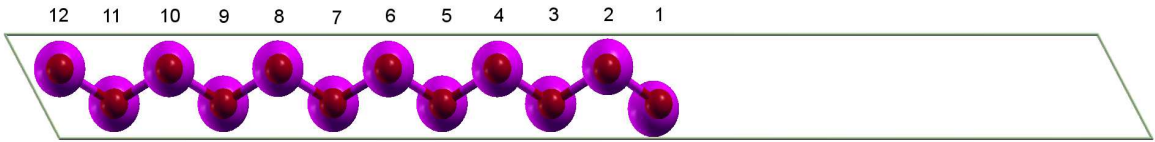}
}
\caption{Relaxed unit cell used to describe the ZnO \{10$\bar{1}$0\} surface: the point of view is along the \textbf{a} axis. The numbers label 
the different atomic layers as used throughout in the manuscript. The large purple balls represent Zn, while the red small ones are for O.}
\label{ZnO_surface_all}
\end{figure}  

The \{10$\bar{1}$0\} is the most stable and frequently occurring ZnO~\cite{wander} surface; it is not polar since it contains both Zn and O atoms. 
The unit cell for the surface is constructed from the primitive cell of bulk ZnO and it is illustrated in figure~\ref{ZnO_surface_all}. Due to the non-polar nature of the surface we found it unnecessary to passivate the side of the slab opposite to the one that gets doped with Co, instead we have chosen to keep the bottom four layers of the slab fixed to the bulk crystal structure during relaxation. Furthermore when performing further relaxation we always constrain the cell 
geometry and allow only the atomic positions to relax. A vacuum region of 10~\AA\ is sufficient to prevent the interaction between slab periodic 
images, and in fact a further increase of the length of the vacuum region produces less than a 1~meV change in the total energy. A slab thickness 
of 17.7~\AA\  (12 atomic layers totaling to 24  atoms) is found to be sufficient to encompass surface relaxations with a less than 5~meV/\AA\ change 
in the maximum force upon introduction of further layers. 

In order to test for surface reconstructions a 2$\times$2 surface cell is created. The atoms, excluding those fixed to their ZnO bulk positions, are randomly 
displaced by a displacement of up to 0.5~\AA. The cell is then allowed to relax, it was found to revert back to the unit cell structure (with no reconstruction).
Therefore, in the absence of any obvious surface reconstruction, we will continue to assume that the primitive surface unit cell can be simply constructed 
from the the primitive ZnO bulk unit cell.
\begin{figure}[ht]
\centering{
\includegraphics[width=0.60\linewidth]{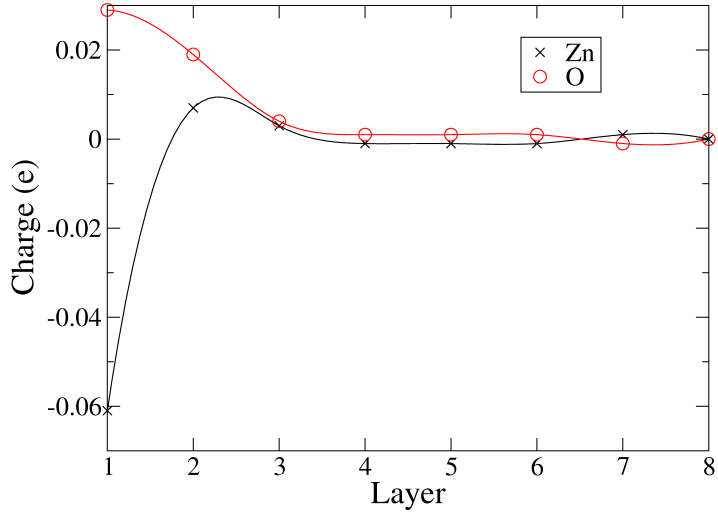}
}
\caption{Change in the M\"ulliken charge, $\Delta\rho$, with respect to the ZnO bulk value in the first atomic layers of the \{10$\bar{1}$0\} surface. The index labeling
the position of the atomic layers follows the definition of Fig.~\ref{ZnO_surface_all}.}
\label{surface_charge}
\end{figure}  

\begin{figure}[ht]
	\centering{
		\subfigure[]{
			\includegraphics[width=0.75\linewidth]{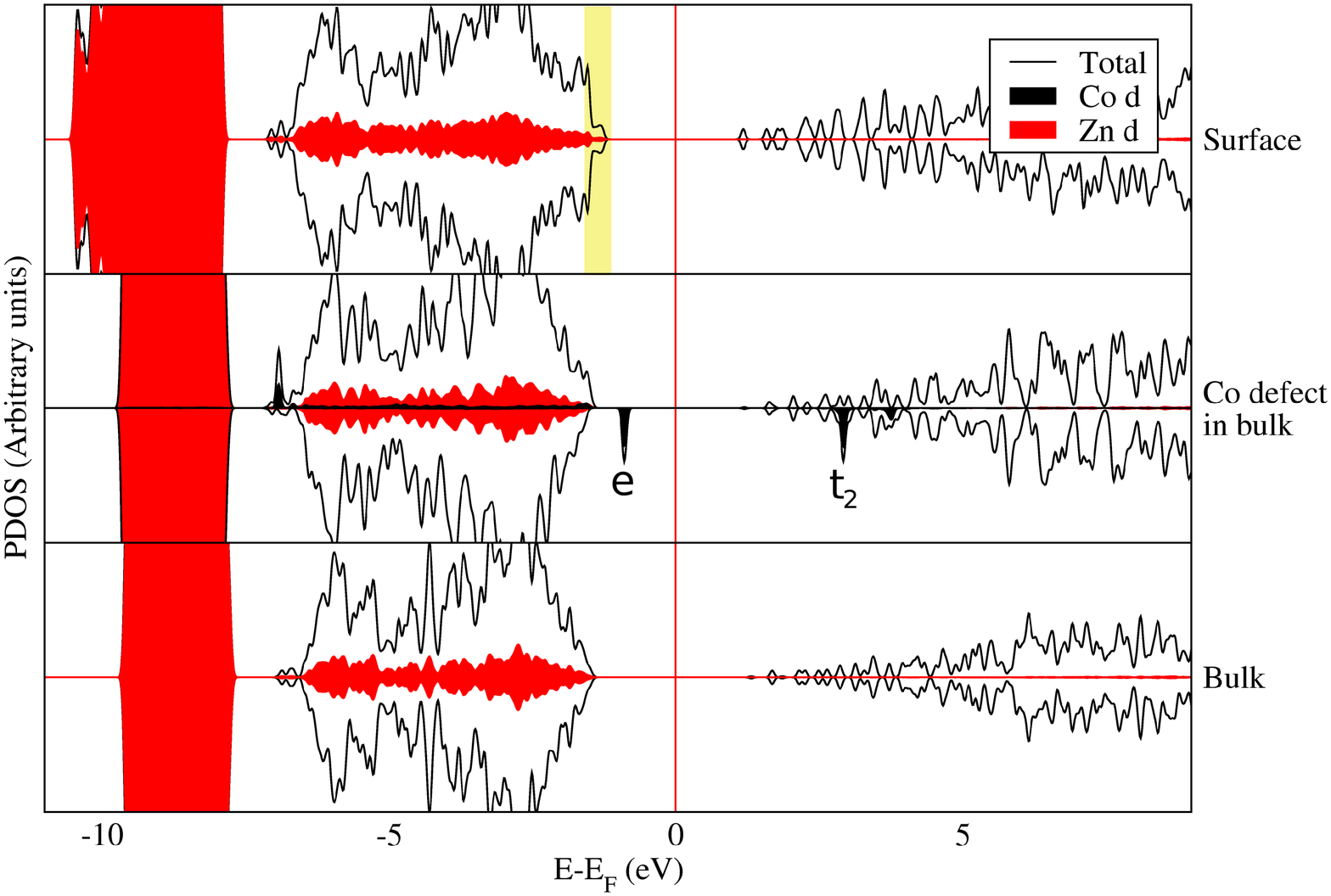}
		\label{pdos}
		}
		\subfigure[]{
			\includegraphics[width=0.15\linewidth]{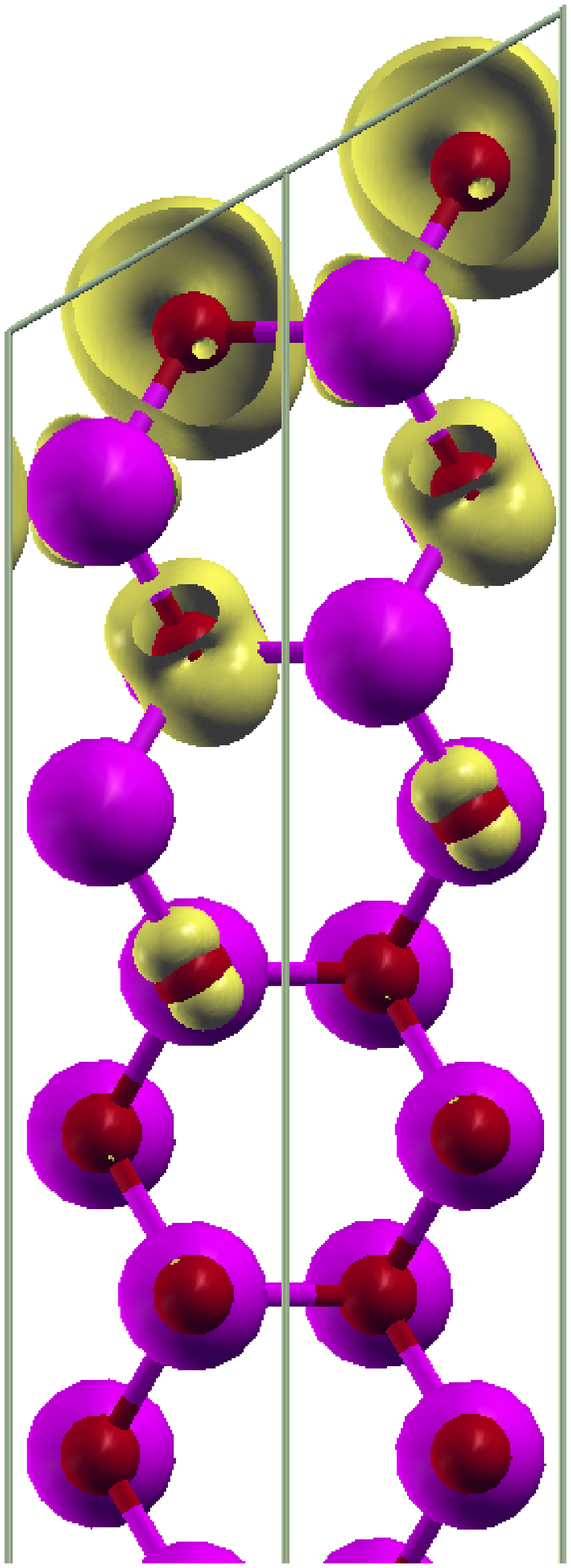}
		\label{surface_state}
		}
	}
\caption{Electronic structure of the ZnO \{10$\bar{1}$0\} surface. In panel (a) we show the DOS for the \{10$\bar{1}$0\} surface (top panel), 
bulk ZnO (bottom panel) and bulk ZnO doped Co at the Zn site (middle panel). The positive DOSs are for the majority electrons while the 
negative ones are for the minority. The energy region marked in yellow in the top panel indicates the energy window used to calculate the 
local DOS of (b). Note the rather localized surface state with amplitude mainly on the O ions. The color code for the atoms in panel (b) is the
same as that of Fig.~\ref{ZnO_surface_all}.}
\end{figure}

In general the formation of the  \{10$\bar{1}$0\} surface reduces the coordination of both O and Zn from 4 to 3 in the top layer, while, of course,
the remaining atoms retain their bulk coordination number. This results in a modification of both the electronic structure and the atomic positions
in the vicinity of the surface. In Fig.~\ref{surface_charge} we present the difference, $\Delta\rho$, between the M\"ulliken charges calculated for the surfaces with
respect to that of bulk ZnO as a function of the layer position (distance from the surface). We can clearly observe that there is a charge re-arrangement
in the topmost three atomic layers. This is more pronounced at the surface (layer 1), where O increases its M\"ulliken valence charge by 0.03~$e$ ($e$ 
is the electron charge), mainly due to the enhanced population of the 2$p$ orbitals. Of course such an enhancement does not correspond 
to doping (the O $p$ shell is already completely filled) but simply to a re-arrangement of the atomic and the overlap M\"ulliken populations, i.e. to
a reduction of the Zn-O hybridization at the surface. In contrast the Zn M\"ulliken charge is reduced by 0.06~$e$, the majority of 
which comes from the 4$p$ orbital. Interestingly the nominal M\"ulliken population for bulk ZnO is already reached after 4 mono-layers for both Zn and O, 
indicating that the perturbation of the electronic structure interests only the surface.

Further information can be obtained by comparing the density of states (DOS) of the surface super cell with that of bulk ZnO [see top panel of Fig.~\ref{pdos}].
Together with an increase of the Zn $d$ DOS bandwidth the main feature of the surface DOS is the presence of a split-off band just at the 
top of the valence band. This is almost entirely due to O $p$ states and it is localized near the surface. In panel Fig.~\ref{surface_state}
we present the isosurface corresponding to the local DOS of such a surface state. This corresponds to the charge density distribution of 
energy levels placed within a narrow energy window around the valence band top [yellow dashed region in Fig.\ref{pdos}]. Clearly the figure
confirms the presence of an O-derived surface state. Finally we note that additional surface states appear at the ZnO conduction band bottom
and, as a consequence, the ZnO band gap is reduced from 3~eV  to 2.5~eV.

\section{Co defects in  ZnO}\label{ZnOCo}

Cobalt ions enter the ZnO lattice as substitutional defects sitting at the Zn site. Their electronic structure and in particular the position of the 
empty $t_2$ states is still a matter of debate \cite{Das_Comment}, although there is agreement that beyond LDA DFT methods are necessary
to capture the insulating nature of ZnO:Co. As mentioned previously here we use ASIC DFT, which returns an electronic structure in good agreement 
with UPS data~\cite{ups}. As an example the DOS of a 128 atoms bulk ZnO cell including one single Co ion is presented in Fig.~\ref{pdos} (middle panel). 
One can clearly see that the filled minority $e$ states lies near the ZnO valence band top, while the first resonant $t_2$ level is well within the
conduction band~\cite{Das_ZnO}. With such an electronic structure $n$-type doping will not open a double exchange channel, while $p$-type 
doping is notoriously difficult in ZnO. Therefore, as it stands bulk ZnO:Co in absence of other defects cannot sustain a long range magnetic interaction.

\begin{figure}[ht]
\centering{
\includegraphics[width=0.90\linewidth]{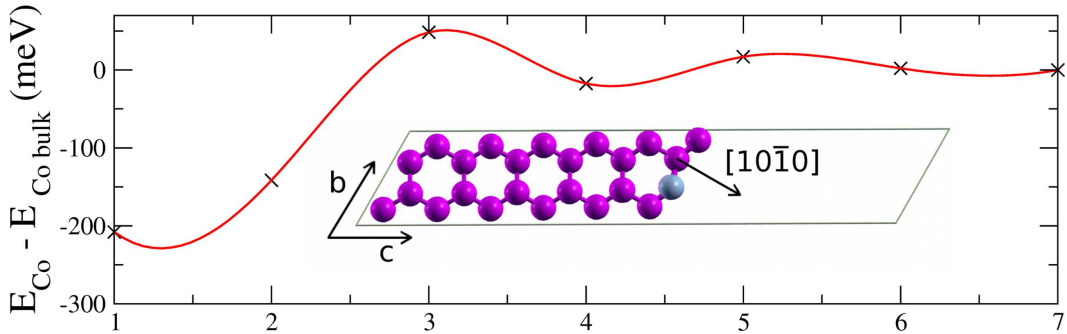}
}
\caption{Energy of incorporating a substitutional Co defect into the ZnO lattice near the \{10$\bar{1}$0\}, calculated for a single Co atom in a 2$\times$2 slab. The surface slab for a substitutonal Co atom in layer 1 is shown in the insert.}
\label{defect_energy}
\end{figure}  
We now move our attention to investigating the effect of the surface on the Co electronic structure and its magnetic interaction. In this case 
a single Co ion is introduced into the slab at different substitutional positions. We use for these calculations a 2$\times$2 slab (comprising 
96 atoms), which corresponds to a total Zn$_{1-x}$Co$_x$O doping of $x=0.21$. Note however the the Co concentration in plane is rather
high, since in a single ZnO atomic layer one out of four Zn is replaced by Co. The cell is always relaxed with the exception of the four bottom
layers, which are kept at their bulk coordinates. In figure~\ref{defect_energy} we plot the energy that a Co ion gains by moving from the bulk
to the surface. This is calculated as the total energy difference between a given super cell with that where Co occupies the middle layer in the
slab. A dipole is observed for these slab calculations, however this dipole remained relatively constant for all the calculations introducing an error of only 5~meV is observed for the worst case, when the Co atom was placed in the top layer. Clearly there is a substantial reduction of the total energy as Co moves closer to the \{10$\bar{1}$0\} surface, indicating that Co segregation to the surface is likely if the ions posses enough kinetic energy during the growth. 

\begin{figure}[ht]
\centering{
\includegraphics[width=0.40\linewidth]{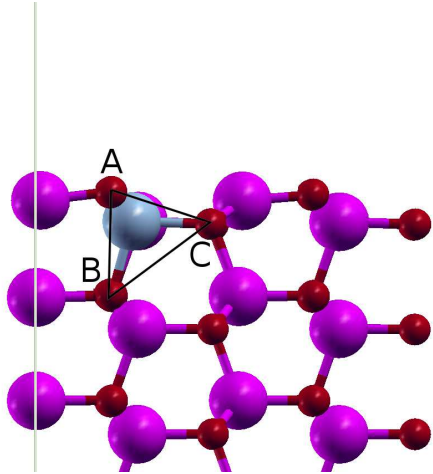}
}
\caption{View along {\bf{b}} showing the distorted oxygen tetrahedron for a Co atom in layer 2.}
\label{5b}
\end{figure}

\begin{figure}[ht]
\centering{
\includegraphics[width=1.00\linewidth]{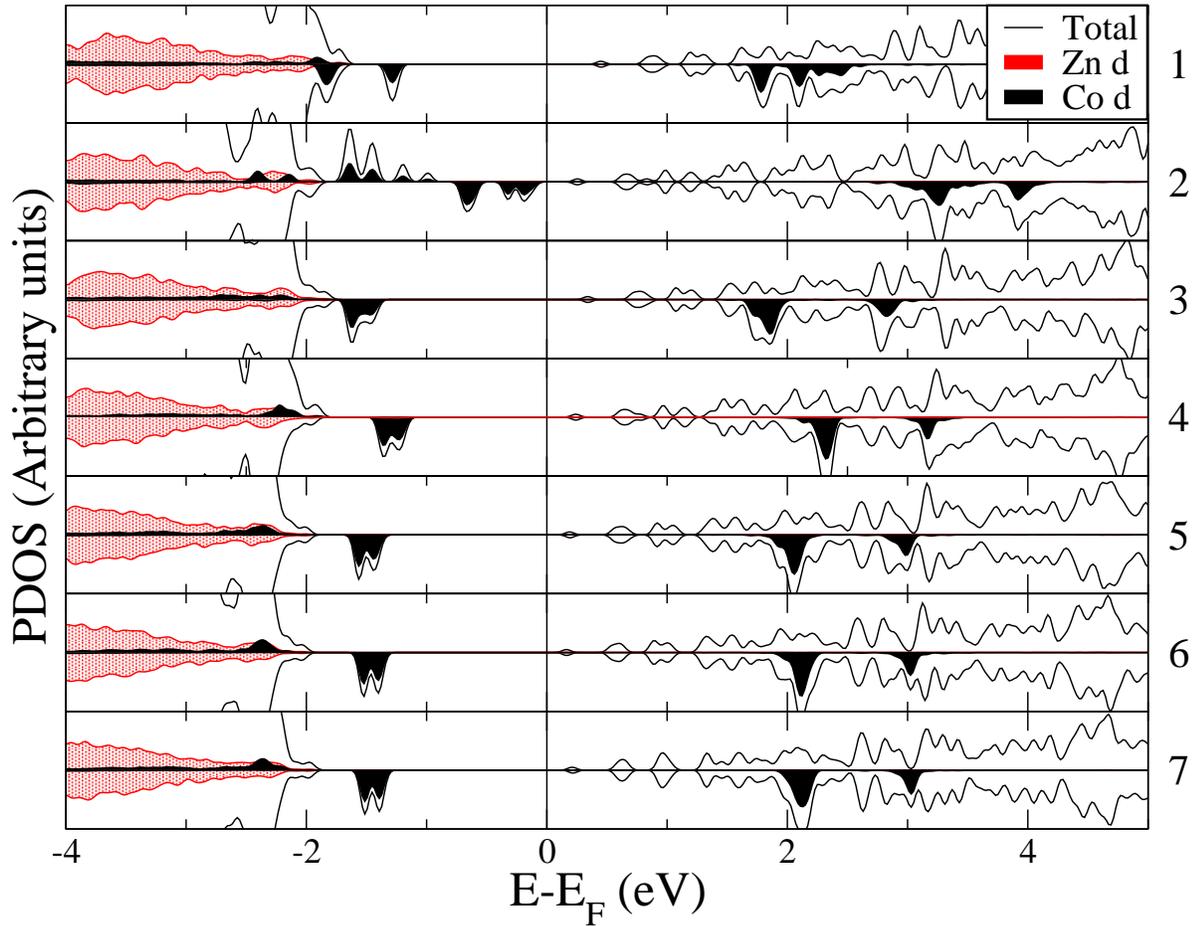}
}
\caption{DOS and orbital projected DOS (PDOS) for a slab including one single Co ion as a function of the distance from the exposed 
\{10$\bar{1}$0\} surface. The number on the right-hand side is the layer index, so that 1 corresponds to Co doped in the first layer,
2 corresponds to Co doped in the second layer and so on. Note that the surface state produce significant changes in the Co DOS
only when Co is placed in the first 2 topmost layers, i.e. at the \{10$\bar{1}$0\} surface.}
\label{pdos_all}
\end{figure}

The electronic structure of Co near the  \{10$\bar{1}$0\} surface is investigated next in figure~\ref{pdos_all}, where we present the DOS of the slab as a function of the Co position. We can observe a substantial interaction between the surface states and the Co $d$ levels, both at the
valence band top and within the conduction band. This is rather strong for the two inequivalent positions at the \{10$\bar{1}$0\} surface
(layer 1 and 2), where a substantial broadening of both the $e$ and $t_2$ peaks is observed. In particular when Co is placed on 
layer 2, which corresponds to the cation site closes to the surface with the full tetrahedral coordination, shown in Figure~\ref{5b}. There is a 
substantial majority $d$ DOS appearing in the ZnO gap and in general all the $d$ manifold is pushed rather high in energy, so that the native 
ZnO band-gap is almost entirely filled. This can be understood from the distortion of the coordinating oxygen tetrahedron. The Co atom relaxes 
to a position closer to the surface, reducing the bond length to the two surface O to 1.89~\AA\ (denoted ``A'' in 
figure~\ref{5b}) and increasing the bond length to the deepest O to 2.00~\AA\ (``B''). The bond length of the remaining O (``C'') 
remains unchanged from the bulk value of 1.96~\AA. This difference in bond length lifts the degeneracy of the $d$ orbitals and results in a 
smeared DOS.  
As the Co moves away from the surface the DOS resembles more closely a superposition between the DOS of the surface and that of Co
doped in the bulk of ZnO. Interestingly for all the positions investigated the orbital degeneracy of minority $e$ states remain lifted
by the interaction with the surface state at the valence band top. Again, with the only exception of Co doping in the two first surface
layers, such an electronic structure is not favorable for sustaining long range magnetic interaction.

\section{Surface mediated magnetic coupling}

We next move on to study the magnetic coupling between Co ions placed in the two topmost layers close to the  \{10$\bar{1}$0\} surface. 
In general we find that the magnetic interaction is strongly dependent on the precise mutual position of the defects and of the free surface.
As such it is necessary to span a large number of possible Co positions. In order to perform this analysis we construct from the relaxed 
super cell discussed in section~\ref{ZnOCo} a 4$\times$3 surface cell including two substitutional Co ions doped in the two topmost layers. 
In order to make the calculations numerically more feasible our constructed cell contains only 6 atomic layers, a number sufficient to keep 
the forces at the surface within 0.01~eV/\AA. This is equivalent to a overall Co concentration of around 4\% which is a value easily achievable 
in experiments (again it should be notice that the concentration of Co is highly inhomogeneous, i.e. it is very large at the surface). 

The exchange coupling is then calculated with the broken symmetry approach, i.e. by comparing the DFT total energy for the same cell,
when the magnetic moments of the two Co ions are aligned either ferromagnetically, $E_\mathrm{FM}$, or antiferromagnetically, $E_\mathrm{AF}$,
with respect to each other. The energy difference $E_\mathrm{AF}-E_\mathrm{FM}$ is then a measure of the magnetic interaction. 
\begin{figure}[ht]
\centering{
\includegraphics[width=0.80\linewidth]{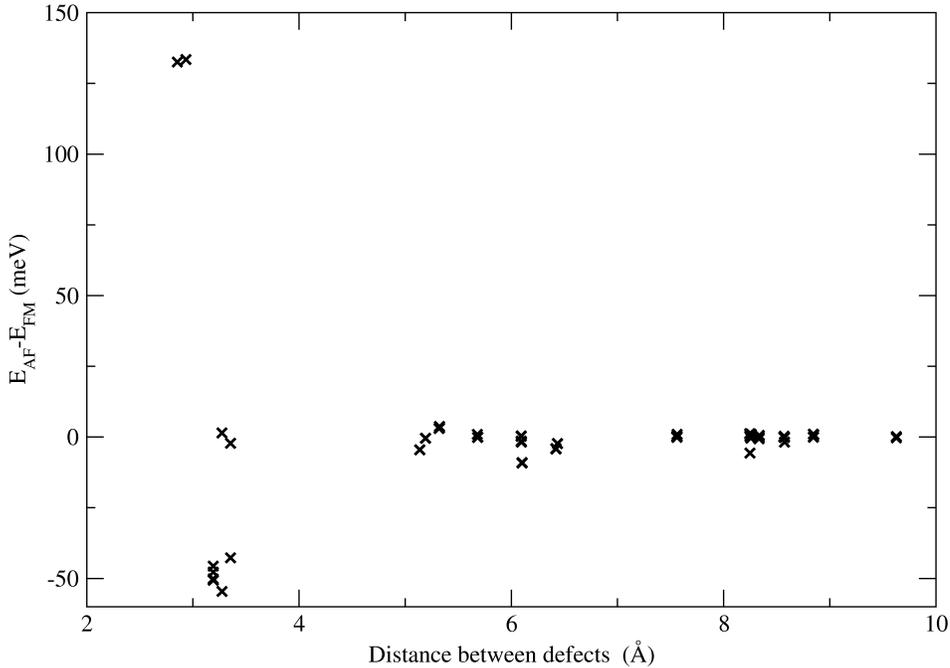}
}
\caption{Difference between the total energies of the ferromagnetic ($E_\mathrm{FM}$) and antiferromagnetic ($E_\mathrm{AF}$) configuration 
of a cell containing two Co ions as a function of the distance between the ions $d_\mathrm{Co-Co}$.}
\label{exchange}
\end{figure}  
The results of this analysis are summarized in Fig.~\ref{exchange}, where $E_\mathrm{AF}-E_\mathrm{FM}$ is plotted against the 
distance between the two Co centers, $d_\mathrm{Co-Co}$. The main features are the presence of two distinct regions of strong magnetic
coupling, respectively ferromagnetic and antiferromagnetic, for short $d_\mathrm{Co-Co}$, and the almost complete absence of long
range magnetic interaction. Thus, unless further doped, the \{10$\bar{1}$0\} surface does not seem to offer a more favorable picture 
in terms of exchange coupling than the bulk, with strong close distance interaction and essentially no long range features. Thus, also
in this case long range ferromagnetism must be excluded. However, whether or not superparamagnetic blocked clusters can form
in abundance and mimic a ferromagnetic hysteresis signal, remains an open question, whose answer sensitively depends on the likely Co
concentration achievable on the surface and the Co magnetic anisotropy. This will be discussed next and put in prospective in the 
closing section.

Going into more details of the magnetic interaction we find that the strong ferromagnetic coupling ($E_\mathrm{AF}-E_\mathrm{FM}>0$) 
is associated to nearest neighbor Co ions, the first being placed right at the surface and the second in the second layer, and exchanged 
coupled through an O also at the surface. In contrast the configurations having strong antiferromagnetic interaction 
($E_\mathrm{AF}-E_\mathrm{FM}<0$) are characterized by having the mediating O atom sitting on the second layer from the surface and
the two Co ions placed either both on the surface or one on the surface and one in the second layer. This suggests that in principle 
specific pattern of magnetic order should be possible if one is able to control the absorption site of the Co ions on the surface.

%    \begin{table}
%        \caption{NN atomic arrangements. The three digit number specifies the atomic configuration for example 112: is the coupling between Co in the first layer via and O in the fist layer with a Co in the second layer.}
%        \centering
%        \begin{tabular}{ll}
%        112, coupling is ferromagnetic with a strength of 132meV&\includegraphics[width=3.5cm]{Fig7a}\\
%        \newline
%        121, coupling is antiferromagnetic with a strength of 50meV&\includegraphics[width=3cm]{Fig7b}\\
%        \newline
%        122, coupling is antiferromagnetic with a strength of 46meV&\includegraphics[width=3.5cm]{Fig7c}\\
%        \newline
%        212, coupling is negligible, calculated to be 1meV&\includegraphics[width=3.5cm]{Fig7d}\\	
%        \end{tabular}
%	\label{table1}
%    \end{table}%

\section{Anisotropy}

In general the breaking of the crystal field symmetry at a surface is expected to increase the magnetic anisotropy, as for instance is the
case of Co on Pt for which the massive zero-field splitting of 9.3~meV has been reported \cite{massive_anisotropy}. For Co doped in 
bulk ZnO the value for the zero-field splitting is known to be 2.76~cm$^{-1}$(0.34~meV) \cite{anisotropy}, with Co having an hard axis
along the wurtzite crystallographic  \textbf{c} axis and an easy plane (the \textbf{a}-\textbf{b} plane). This is consistent with model Hamiltonian
calculations for transition metals in trigonally distorted tetrahedral coordination~\cite{Kuzian}. With such a value in hand and assuming
a certain degree of frustration we have previously estimated that superparamagnetic particles comprising of more than 250 Co ions should 
be blocked at room temperature \cite{Das_ZnO}. The question is now whether magneto-crystal anisotropy at the surface can push such size 
limit towards smaller particles.

In order to calculate the magneto-crystal anisotropy we use a 2$\times$2 surface cell containing 4 atomic layers for a total of 32 atoms, and where
one surface Zn ion is replaced by Co. An equivalent
cell, although with periodic boundary conditions along the three dimensions, is used for the bulk calculations. The anisotropy is computed from 
{\it Siesta} DFT by using an on-site approximation for the spin-orbit matrix elements \cite{miguel}. This requires a non-collinear calculation, which unfortunately
cannot be performed with the ASIC method, since at present it is only defined for collinear DFT. Therefore all the anisotropy calculations are 
carried out at the LDA level by using the Ceperley-Alder~\cite{Ceperley80} functional. In practice we compute the total energy as a function of 
the direction of the Co magnetic moment with respect to the wurtzite \textbf{c} axis.

\begin{figure}[ht]
	\centering{
		\subfigure[]{
			\includegraphics[width=0.47\linewidth]{Fig8}
		\label{anisotropy}
		}
		\subfigure[]{
			\includegraphics[width=0.47\linewidth]{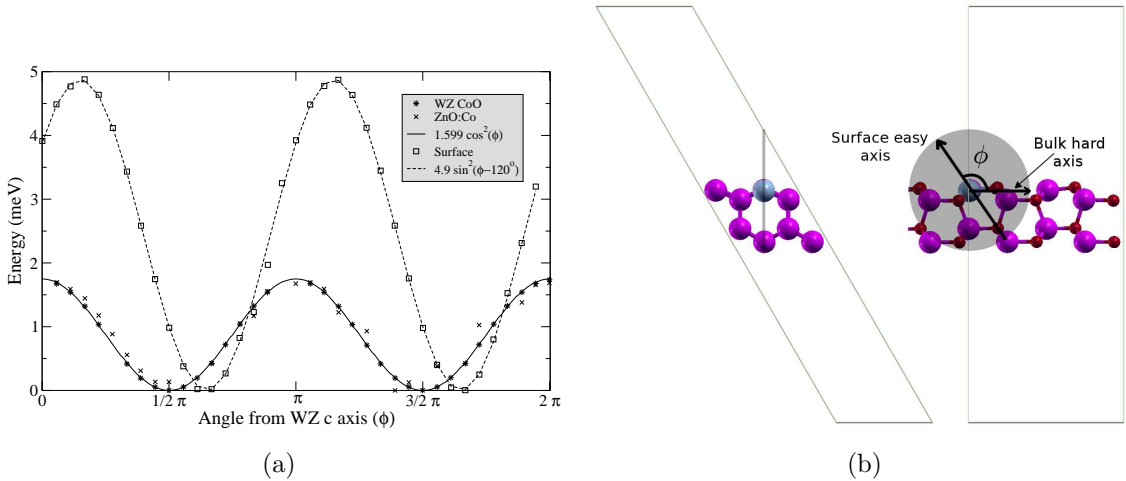}
		\label{anisotropy_insert}
		}
	}
\caption{(a) Total energy, $E$, as a function of the angle $\phi$, between the Co magnetic moment and the wurtzite \textbf{c} axis. 
The calculations are performed for Co in a bulk ZnO matrix (ZnO:Co), for wurtzite CoO (WZ CoO) and for a Co ion at the 
\{10$\bar{1}$0\} ZnO surface (surface). (b) shows the path 
taken by the magnetic moment for the surface cell. Note $\phi$=0 corresponds to the bulk \textbf{c} direction.}
\end{figure}

The results of our calculations are presented in figure \ref{anisotropy}, where we report the total energy, $E$, as a function of the angle, $\phi$, between
the magnetization and the wurtzite \textbf{c} axis. In the figure we present three cases, namely that of Co in bulk ZnO, of wurtzite CoO \cite{Ruairi2} and
the Co doped \{10$\bar{1}$0\} surface. For both the bulk ZnO:Co and wurtzite CoO, we predict a hard axis along the crystallographic wurtzite \textbf{c} direction, and an easy plane perpendicular to this axis, in agreement with experiments. The zero-field splitting parameter is estimated by a simply fit to our DFT calculations to be 0.7~meV (assuming $S=3/2$ for Co) and it is essentially identical in the two cases. This is approximately a factor two larger than what is found in experiment. We are at present uncertain of the reasons for such a discrepancy. This may lie in the choice of the LDA functional, which usually 
over-estimates the $p$-$d$ hybridization, or in the on-site approximation to the spin-orbit matrix elements, or in the crude way we use to extract the parameter. Even with this uncertainly it is still useful to compare the bulk anisotropy to that of Co on the surface. 

We find that the surface changes drastically the nature of the anisotropy. A Co ions placed at the under-coordinated position on the surface 
now has an easy-axis oriented at $120^\circ$ from the wurtzite \textbf{c} axis as shown in Figure~\ref{anisotropy_insert}. The zero field splitting 
is calculated 1.97~meV, i.e. it is approximately a factor three larger than that in the bulk. Assuming that the factor 2 error on our 
estimate is similar for both bulk and the surface, this still gives us a quite remarkably high anisotropy, which pushes the limit for 
superparamagnetic clusters blocked at room temperature to about 80 atoms. This means that we need approximately 80 Co atoms,
magnetically coupled (i.e. in nearest neighbor positions) on the surface, in order to have a cluster displaying magnetic hysteresis at room temperature. 

\section{Concluding Remarks}

We have studied the electronic structure, magnetic interaction and magnetic anisotropy of Co ions at or in the vicinity of the \{10$\bar{1}$0\} 
ZnO surface, in the search for an explanation of the often claimed room temperature ferromagnetism in ZnO:Co. Our results give a
contrasting picture. On the one hand we have found that the magnetic interaction is rather strong for atoms that sit on the surface, which is in some circumstances ferromagnetic in nature. Moreover there is a substantial energy gain for a single Co ion to locate at the
surface, so that high superficial Co densities might be expected in nanoparticles and granular materials, i.e. in objects with a large
surface to volume ratio. On the other hand we also find that the magnetic interaction remains extremely short-ranged, which means
that the surface does not act as an extended defect mediating the magnetic coupling. This aspect is not different from what happens
in bulk ZnO and excludes the presence of long-range room temperature ferromagnetism. Certainly the situation might change in the
case of polar surfaces or when surface doping is possible. 

There are however a few more facts to consider. Firstly the magnetic anisotropy at the \{10$\bar{1}$0\} surface is threefold enhanced 
with respect to its bulk value and it changes from hard-axis easy-plane to easy-axis. This means that superparamagnetic clusters formed 
at the surface will block at much higher temperature than the in bulk. Alternatively one can say that for the same blocking temperature, 
much smaller clusters are required at the surface to mimic a ferromagnetic signal. This is indeed an attractive prospective which may 
validate the uncompensated spin model proposed by Dietl et al.~\cite{ditel}. 
Using a rather crude estimate of the size of superparamagnetically blocked clusters at room temperature we predict that approximately 
80 atoms are required. This means that one needs up to 80 atoms connected by a nearest neighbor percolation path on the \{10$\bar{1}$0\} 
surface. Considering that the two dimensional percolation threshold ranges from 1/2 to about 0.7 depending on the particular surface, 
we can conclude that the majority of the surface Zn sites should be replaced by Co in order to have magnetic hysteresis at room temperature.

This work is sponsored by Science Foundation Ireland and the EU FP7 project ATHENA. We thank HPC-Europa and TCHPC for computational 
support. \\

\bibliographystyle{plain}

\end{document}